\documentclass[12pt]{article}
\usepackage[T1]{fontenc}
\usepackage{times}
\begin{document}

\title{Collapse of a randomly forced particle: a comment}

\author{Stephen J. Cornell$^1$, Michael R. Swift$^2$, and Alan J. Bray$^3$\\
\\$^1$Department of Zoology, University of Cambridge, \\Cambridge CB2 3EJ, UK.
\\$^2$School of Physics and Astronomy, University of Nottingham, \\Nottingham NG7 2RD, UK.
\\$^3$Department of Physics and Astronomy, University of Manchester, \\Manchester M13 9PL,
UK.}

\maketitle
\begin{abstract}
We refute the arguments by Anton in cond-mat/0004390, which set out to disprove
the existence of a collapse transition for a randomly forced inelastic particle.
\end{abstract}

An eprint by Anton \cite{anton} has recently appeared on this archive, criticizing
our proposal of a collapse transition for an inelastic, randomly-forced particle
\cite{nutter}. Our conclusions were that a particle forced by Gaussian white
noise that rebounds from a wall with coefficient of restitution \( r
\) will, with probability 1, dissipate all its energy and come to rest
at the wall after an infinite number of collisions in a finite time,
provided \( r<r_{c}=e^{-\pi /\sqrt{3}} \).  It has been pointed out
that the transition is not present for certain discretizations of the
Langevin equation \cite{bonfim}, and it is also not clear how it
manifests itself in experimentally realizable systems
\cite{nutter2}. However, ref.~\cite{anton} goes further, and argues
that the transition is also absent for the ideal case of a pure
white-noise force in continuous time. The purpose of this Comment is
to point out substantial errors in Anton's arguments, which, we
believe, invalidate his conclusions.

The existence of the collapse transition is supported by analytical calculations
\cite{burk} by Burk\-hardt, Franklin, and Gawronski (henceforth BFG), who constructed
a steady-state solution of the Fokker-Planck equation for a particle
confined in a finite spatial interval. They observed that the solution
is well-behaved when the coefficient of restitution \( r \) is in the
range \( r_{c}<r<1, \) where \( r_{c} \) is the critical value for
collapse proposed in \cite{nutter}, whereas the rate of collisions of
the particle with the boundary diverges as
\( r \) approaches \( r_{c} \) from above. Anton's criticisms of this work,
together with our responses, are as follows:

\begin{itemize}
\item \emph{Burkhardt et al.'s probability density function is not normalizable}.
This is incorrect. BFG state explicitly that their solution is normalizable
for \( r_{c}<r<1 \), and it is in fact the onset of non-normalizability as
\( r\to r_{c}^{+} \) that lead them to support our conclusions. 
\item \emph{The system does not approach a steady state, 
since the moments of a time-dependent
solution are divergent in time}. Anton bases this on two
calculations. The first shows that the energy of an \emph{elastic}
particle (\( r=1 \)) diverges linearly in time, a result which is also
stated in ref.~\cite{burk}. The second is a calculation (incorrect, as
we later shall argue) of spatial moments for an inelastic particle (\(
r<1 \)) in a \emph{semi-infinite} region, which diverge in time.
Neither of these calculations are relevant to the case of an
\emph{inelastic} particle in a \emph{finite} region, as studied by
BFG. On the other hand, physical considerations show that these
moments cannot diverge in that case. Firstly, the spatial moments are
restricted due to the spatial confinement of the particle.  Secondly,
the energy is prevented from diverging by the fact that, at high speed
\( v \), energy is dissipated at a rate \( \propto (1-r^{2})v^{3} \) (collision
rate \( \propto v \), energy dissipation per collision \( =(1-r^{2})v^{2} \)),
whereas energy is being supplied at a constant rate \cite{foot2}.
\item \emph{The boundary conditions used by BFG lead to a contradiction}. \emph{}The
argument is as follows. The boundary condition for the probability density function
\( G(x,v) \) used by BFG, which corresponds to the stipulation that particles
incident on the boundary at velocity \( -v \) rebound at \( rv \), is
\begin{equation}
\label{eq1}
vG(0,-v)dv=(rv)G(0,rv)d(rv).
\end{equation}
Setting \( v=1 \), and then replacing \textbf{\( r\to w \)} leads to 
\begin{equation}
\label{eq2}
G(0,w)=\frac{1}{w^{2}}G(0,-1).
\end{equation}
However, contrary to Anton's claims, this does not imply that the flux
\( vG(0,v) \) has a non-integrable singularity at \( v=0 \). The
reason is that the solution must depend upon the coefficient of
restitution---i.e., \( G \) has a parametric dependence upon \( r \),
which has been suppressed in the above notation. This means that, for
a given value of \( r \), eqn.~(\ref{eq2}) is only valid for a single
value (\( =r \)) of \( w \). If we include explicitly the \( r
\)-dependence in \( G(x,v;r) \), eqn~(\ref{eq2}) becomes
\[
G\left( 0,w;w\right) =\frac{1}{w^{2}}G\left( 0,-1;w\right) ,\]
which does not imply that there is a non-integrable singularity in \( vG(0,v,r) \)
at constant~\( r \).
\end{itemize}
The final point leads Anton to propose alternative boundary
conditions, which contradict eqn.~(\ref{eq1}) above. However, we have
shown that there is no reason for believing eqn.~(\ref{eq1}) to be
wrong, whereas Anton's alternative is not justified on physical
grounds. We therefore believe that he has not constructed a valid
solution for this system, and the ensuing conclusions in sections III
and IV of ref.~\cite{anton} are erroneous.

Ref.~\cite{anton} contains a further, independent argument against the
existence of collapse. Anton states that, when collisions with the
boundary occur, the Langevin equation for a damped particle near an
elastic wall reduces to the equation of motion for the inelastic,
undamped particle. Comparing the discretized versions of these
equations at the instant of collision with the boundary \cite{foot3},
we find
\begin{eqnarray*}
\hbox {Damped,\, elastic:}\qquad v(t+\Delta t) & = 
& -v(t)-\Delta t\left\{ \eta _{1}(t)-(1-r)v(t)\right\} \\
\hbox {Undamped,\, inelastic:}\qquad v(t+\Delta t) & = 
& -rv(t)-\Delta tr\eta (t)
\end{eqnarray*}
These are far from equivalent---the inelastic particle loses a finite
fraction of its energy at the instant of collision, whereas the damped
particle loses an infinitesimal fraction of its energy. The equations
become the same if one sets \( \Delta t=1 \), but this does not allow
for a systematic approach to the continuum limit since the Langevin
equation has already been adimensionalized (i.e., the timescale has
already been set in eqn.~(6) of ref.~\cite{anton}).  In any case, it
is already known \cite{bonfim,nutter2} that inelastic collapse is
destroyed by such crude discretization schemes.

To summarize, we do not believe that Anton has provided a substantive
argument against the existence of the collapse transition. However, we
agree that further work is needed to investigate whether, in the ideal
system, inelastic collapse is definitive (i.e., is a particle that
comes to rest at the wall capable of escaping again in a finite
time?), and also what remnant of the collapse transition occurs in
real systems.

We would like to thank Lucian Anton for interesting discussions.


\begin{thebibliography}{1}
\bibitem{foot2}The velocity performs a random walk, so the set of trajectories where \( v^{2} \)
increases as \( t^{\mu } \), with \( \mu >1 \), is of measure zero.
\bibitem{foot3}Equation (6) in \cite{anton} apparently contains a misprint, and should read
\( \ddot{x}=-(1-r)\dot{x}+\eta _{1}(t) \).
\end{thebibliography}
\end{document}